\documentclass[aps]{revtex4} 
\usepackage{psfig}
\usepackage{epsfig}
\usepackage{bm}

\newcommand{\be}{\begin{equation}}
\newcommand{\ee}{\end{equation}}
\newcommand{\bv}{{\bm v}}

\begin{document}

\title{A molecular dynamics ``Maxwell Demon'' experiment for granular 
mixtures\footnote{Paper submitted to the special issue of Molecular Physics
edited by J.-P. Hansen and R. Lynden-Bell on the occasion of Dominique 
Levesque's 65th birthday, }}
 
\author{Alain Barrat
and Emmanuel Trizac
}

\affiliation{
Laboratoire de Physique Th{\'e}orique
(UMR 8627 du CNRS), B{\^a}timent 210, Universit{\'e} de
Paris-Sud, 91405 Orsay Cedex, France 
}
 
\date{\today}

\begin{abstract}
We report a series of molecular dynamics simulations and investigate
the possibility to separate a granular mixture of inelastic hard
spheres by vigorously shaking it in a box made of two connected
compartments.  As its one-component counterpart, the system exhibits a
``left-right'' symmetry breaking entirely due to the inelasticity of
grain-grain collisions, and triggered by increasing the number of
particles. In the compartment where the density of grains is larger,
we observe a partial segregation with a predominance of heavy
particles. However, this compartment still has a higher density of
light particles than the other one, which is light-rich. The density,
granular temperature and anisotropic pressure profiles are
monitored. We also discuss how to construct a relevant order parameter
for this transition and show that the resulting bifurcation diagram is
dominated by large fluctuations.
\end{abstract}

\maketitle 

\section{Introduction}

Although granular matter may exhibit similarities with molecular
fluids (such as pattern formation), it is nevertheless intrinsically
out of equilibrium: The inter-particle collisions dissipate kinetic
energy, and a steady state may only be achieved by a suitable energy
supply. As a result, such systems may display many phenomena that are
``forbidden'' by the laws of equilibrium statistical mechanics. In the
realm of granular gases (dilute systems of macroscopic grains in rapid
motion and colliding inelastically), tendency to form 
clusters~\cite{Kada,Gold,McNa,EPJE}, non Gaussian velocity 
distributions~\cite{Losert,Rouyer,Noije,Puglisi,Nie,Cafiero,JPA,Prevost,vibrated}, 
long range velocity 
correlations~\cite{Prevost,Pre_twan,Blair,Moon,Pre2}, and breakdown of kinetic
energy equipartition in a mixture of dissimilar 
grains~\cite{Garzo,Feitosa,Wildman,Granular,vibrated} have been reported.

Another interesting feature, at complete variance with equilibrium
phenomenology has been obtained with a simple 
experiment~\cite{Schlichting,Weele,Meer,Mikk}: a vibrated system of grains
confined in a box with two connected identical compartments may
exhibit a stationary state with spontaneous symmetry breaking
(non-equipartition of grains between the two compartments). 
This clustering phenomenon may be
interpreted as a separation in a ``hot'' and a ``cold'' region,
considering that the granular temperature is a direct measure of the
mean squared velocity of the particles. In the limit where the
exchange of particles between the two compartments may be considered
as an effusion process, Eggers~\cite{Eggers} has put forward an
analytical approach to explain this apparent intrusion of a ``Maxwell
Demon''. On the other hand, Brey and collaborators~\cite{Brey_Maxwell}
reported a hydrodynamic mechanism for the symmetry breaking, that
becomes operational under some simplifying assumptions in the opposite
limit where the size of opening connecting the two compartments is larger
than the mean-free-path of the gas in its vicinity.

In this contribution, we revisit numerically the Maxwell Demon 
experiment in the latter case,
and consider the specific situation of a binary low density
granular mixture, with the aim to investigate whether such a set-up is
able to achieve an efficient segregation of the mixture. The model is
defined in section~\ref{sec:model}.  Making use of molecular dynamics
simulations, we discuss in section~\ref{sec:bifurc} how to construct a
relevant order parameter for the transition under study, and show that
it is dominated by large fluctuations (as also observed recently in a
related context~\cite{Meerson}).  The two components of the mixture
are found to behave differently: heavy particles display a stronger
(left-right) asymmetry than the light ones, leading to a separation
between a dense gas rich in heavy particles and a dilute light-rich
gas. The behaviour of partial densities and granular temperatures are
investigated (section~\ref{sec:density}) from which we deduce 
the different components of
the pressure tensor making use of the general equation of
state derived in~\cite{vibrated}; although the hypothesis of an
isotropic pressure given by the ideal gas equation of state is clearly
not verified, we show that the no-convection hydrodynamic condition
of a divergence-free pressure tensor ($\nabla \cdot P=0$) 
is obeyed, taking into account anisotropies,
boundaries and corrections to the ideal gas equation of
state. Conclusions are finally drawn in section~\ref{sec:concl}.

\section{The model}
\label{sec:model}

The system is made of $N$ inelastic hard disks evolving in a
$S\times L$ two-dimensional box, losing energy at inter-particle
collisions and gaining energy through collisions with two vibrating
walls situated at $y=0$ and $y=L$.  The particles have diameters
$\sigma_i$ and masses $m_i$, $i=1,2$.  A binary collision between
grains of species $i$ and $j$ is momentum conserving and dissipates
kinetic energy: the collision $i$-$j$ is characterized by the
coefficient of normal restitution $\alpha_{ij}$.  Accordingly, the
pre-collisional velocities (${\bm v}_i, \bv_j$) are transformed into
the post-collisional couple (${\bm v}'_i, {\bm v}'_j$) such that
\begin{eqnarray}
\bm{v}_i' \, =\,  \bm{v}_i - \frac{m_j}{m_i+m_j} (1+\alpha_{ij})
(\widehat{\bm{\sigma}}\cdot \bm{v}_{ij})\widehat{\bm{\sigma}} 
\label{eq:coll1}\\
\bm{v}_j'\, = \, \bm{v}_j + \frac{m_i}{m_i+m_j} (1+\alpha_{ij})
(\widehat{\bm{\sigma}}\cdot \bm{v}_{ij})\widehat{\bm{\sigma}}
\label{eq:coll2}
\end{eqnarray}  
where $\bv_{ij}=\bv_i-\bv_j$ and $\widehat{\bm\sigma}$ is the center
to center unit vector from particle $i$ to $j$. Note that
$\alpha_{ij}=\alpha_{ji}$ to ensure the conservation of total linear
momentum $m_i \bv_i+m_j\bv_j$. The total density is denoted $\rho$,
and the partial densities $\rho_i = x_i \rho$ (the number fractions
$x_i$ are such that $\sum_i x_i=1$). The granular temperature of species $i$
is $T_i$, defined from the mean kinetic energy of subpopulation $i$, by 
analogy with the usual temperature of elastic gases:
$T_i = \langle m_i v_i^2\rangle/d$, where $d$ is the space dimension
(here $d=2$). In the remaining of the paper this granular temperature will
be coined ``temperature''for simplicity.

The box is divided into two compartments of width $S/2$ by a wall
parallel to $Oy$ starting at height $y_0$.  The walls located at $x=0$
and $x=S$ are elastic, while those at $y=0$ and $y=L$ are vibrating
and thus inject energy into the system.
For simplicity, the two vibrating walls are taken to move in a
saw-tooth manner, so that a colliding particle at $y=0$ (resp. $y=L$)
always finds the wall to move ``upwards'' (resp. ``downwards'') with
the same velocity $v_0$ (resp $-v_0$).  In addition, the amplitude of
the vibration is considered to vanish (i.e. to be much smaller than
the local mean free path~\cite{Eggers,Brey_Maxwell}), so that the 
walls are located at
the fixed positions $y=0$ and $y=L$: the $y$-component velocity of a particle
colliding with the wall at $y=0$ (resp. $y=L$) is therefore changed
according to $v'_y = 2v_0-v_y$ (resp. $v'_y = -2v_0-v_y$). Since we
consider vigorous shakings, the gravitational field has not been
included in the analysis.

 For simplicity, we have considered equimolar mixtures ($N_1=N_2$) of
particles having the same diameter ($\sigma_1=\sigma_2$) but different
masses. Various mass ratios $m_1/m_2 \in [1:5] $ have been
studied~\cite{Rqe}, so that the species $1$ is always the heavier
particle. We have run molecular dynamics simulations~\cite{Allen}
changing $N$ either at constant packing fraction (equal to $\pi \rho
\sigma^2/4$ in two dimensions) or at constant $\sigma_i$, with the
same qualitative observations.  The numerical results we will present
correspond to a fixed low mean packing fraction $\eta_0=0.015$ (the
inelastic collapse~\cite{McNa} occurring if the mean density exceeds a
low threshold), and to equal coefficients of restitution
$\alpha_{ij}=0.9$, close to experimentally relevant values. We have
investigated other values of the restitution coefficients between
$0.7$ and $0.9$, and two different aspect ratios, $L=S$ and $L=2S$,
with the same qualitative results.

\section{Bifurcation diagram and large fluctuations}
\label{sec:bifurc}

For a one component system, it has been shown from a hydrodynamic
approach~\cite{Brey_Maxwell} that, as the number of
particles in the box ($N$) is increased, a transition occurs
at a certain threshold $N^*$: for $N < N^*$, the system is symmetric, i.e
the mean number of particles in each compartment is $N/2$ while,
for $N > N^*$, one of the compartments becomes 
more populated and colder than the other. This hydrodynamic study
relies of the assumption of an isotropic pressure given by the ideal gas
equation of state~\cite{Brey_Maxwell}.
At a given inelasticity (i.e. at given values of the restitution
coefficients), the control parameter (governing the transition
from symmetric to asymmetric situation) is proportional
to $N \sigma^{d-1} /S$, where $\sigma$ is the particle 
diameter~\cite{Brey_Maxwell}.  At fixed reduced density $n=N \sigma^d/(LS)$ 
the above parameter is proportional to $N^{1/d}$, while at fixed size
$\sigma$ it scales like $N$.

\begin{center}
\begin{figure}[ht]
\centerline{
\epsfig{figure=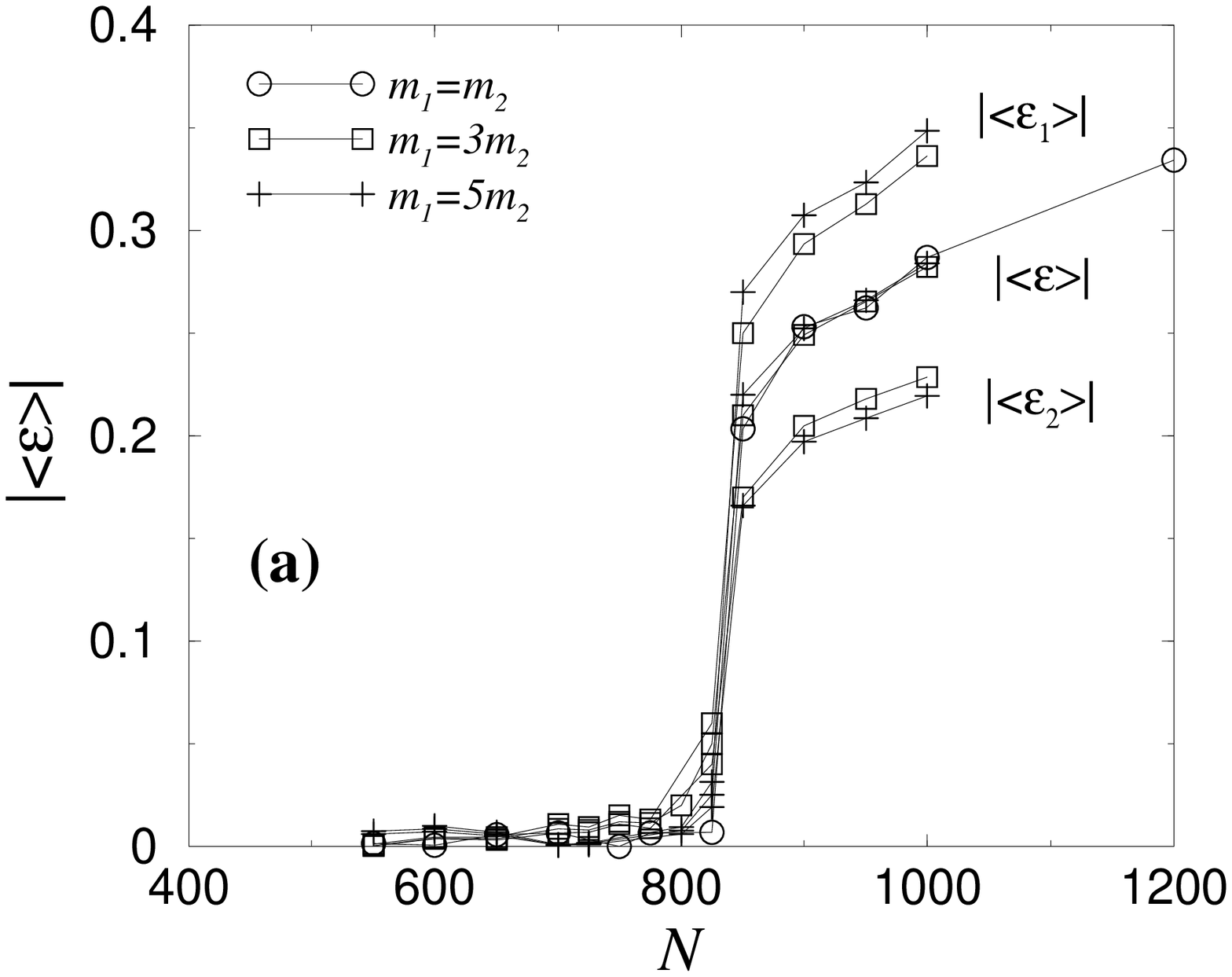,width=8cm,angle=0}
\epsfig{figure=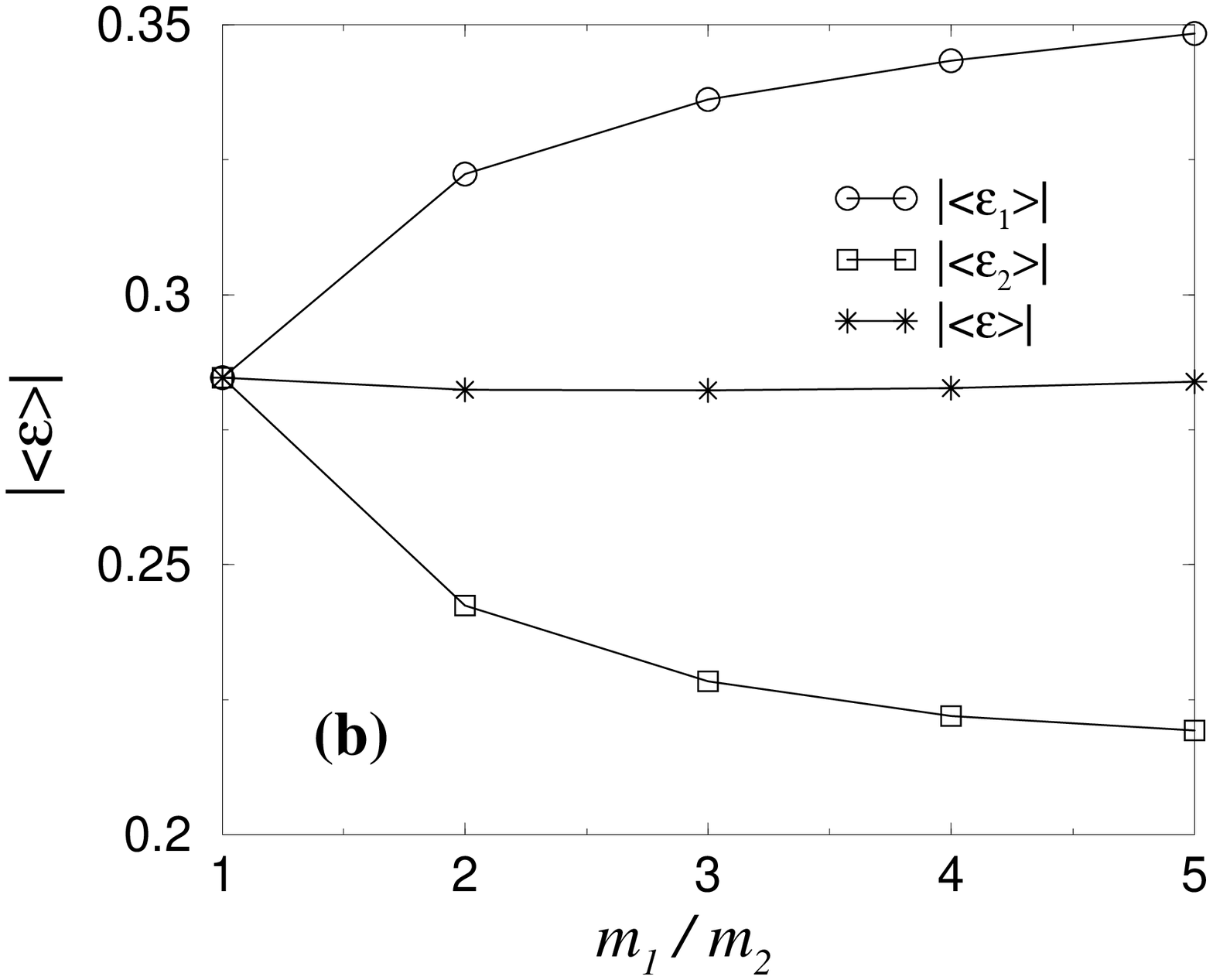,width=8cm,angle=0}
}
\caption{
(a): Asymmetry parameters $|\langle\epsilon\rangle|$ and
$|\langle\epsilon_i\rangle|$ versus number of particles for three
different mass ratios and a given simulation time, i.e. number of collisions
per particle. 
Here, all the inelasticity parameters are taken
equal: $\alpha_{11}=\alpha_{12}=\alpha_{22}=0.9$.  The opening connecting
both compartments is 40\% of the total height of the simulation cell
($y_0=0.4 L$).\\ (b): Asymmetry parameters versus mass ratio, at fixed
number of particles $N=1000$ and $\alpha_{ij}=0.9$.  As the mass ratio
increases the asymmetry increases for the heavy particles and
decreases for the light ones. As emphasized in the text, these figures
depend on the simulation time available. }
\label{fig:diagram}
\end{figure}
\end{center}

The ``order parameter'' of this transition was 
defined in~\cite{Eggers,Brey_Maxwell} as the time average
$\langle \epsilon\rangle$ of the asymmetry $\epsilon$:
\begin{equation}
\epsilon = \frac{ N - 2  N^{left}  }{2N}
\end{equation}
where $ N^{left}$ is the number of particles in the left
compartment. In addition to the global $\epsilon$, we may introduce
two relevant asymmetry parameters for each type of particles:
\begin{equation}
\epsilon_i = \frac{ N_i - 2  N_i^{left}  }{2N_i}, \ \ i=1,2.
\end{equation}
{\em For a given simulation time}, if one computes $\langle
\epsilon_i\rangle$ for the binary mixture (or $|\langle
\epsilon_i\rangle|$ to have a positive quantity), a left-right
symmetry breakdown is evidenced (see Fig.~\ref{fig:diagram}). The
asymmetry is more pronounced for heavy particles 
($| \langle \epsilon_1 \rangle | >  | \langle \epsilon_2 \rangle |$), 
and $| \langle \epsilon_1 \rangle |$ increases with the mass ratio
$m_1/m_2$. On the other hand, the light particles asymmetry decreases
with $m_1/m_2$.  At this point we conclude that the compartment with
larger global density is heavy-rich, while the lighter particles are
more uniformly distributed and therefore the less populated
compartment is richer in light particles.

\begin{center}
\begin{figure}[ht]
\centerline{
\epsfig{figure=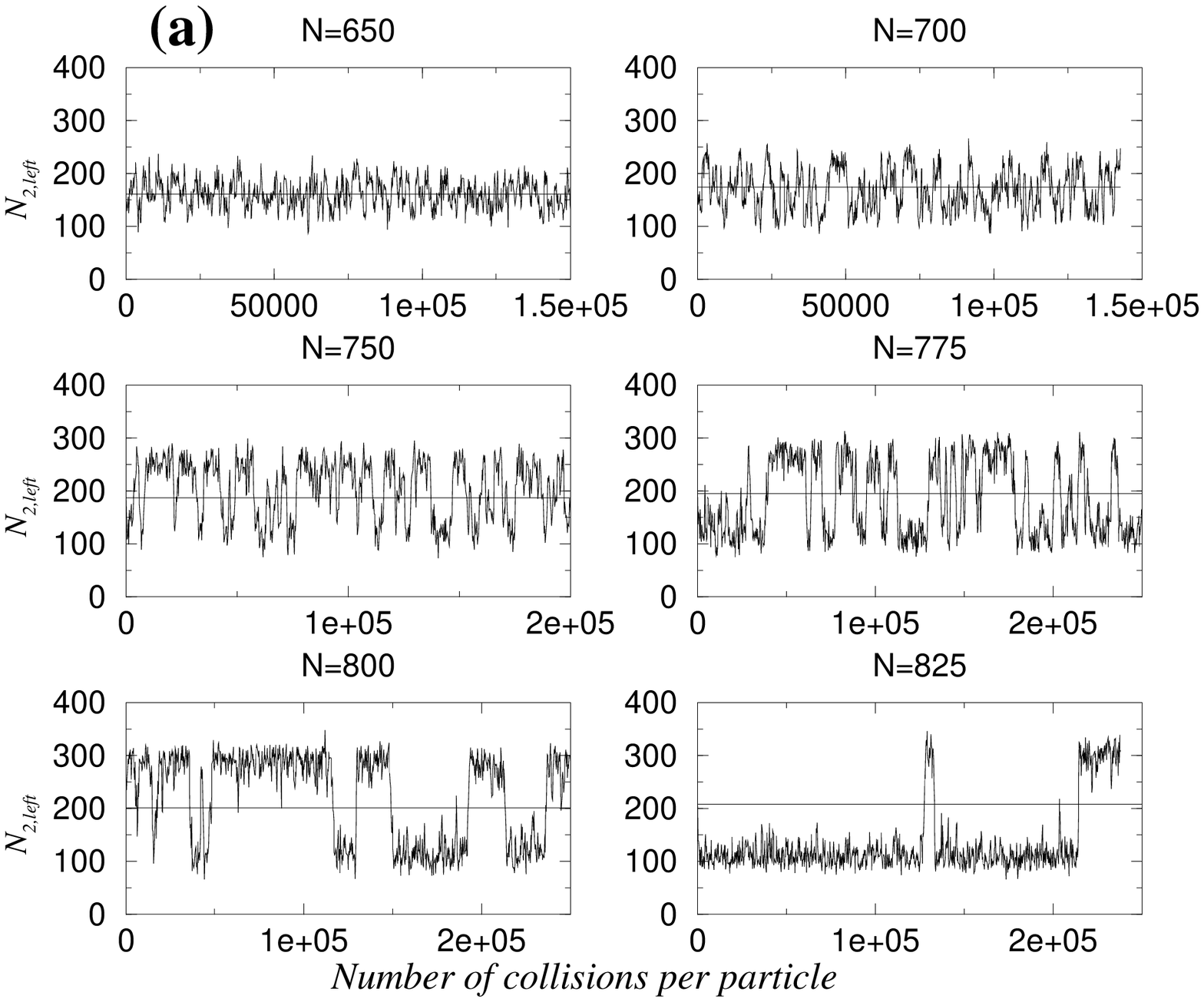,width=8cm,angle=0}
\epsfig{figure=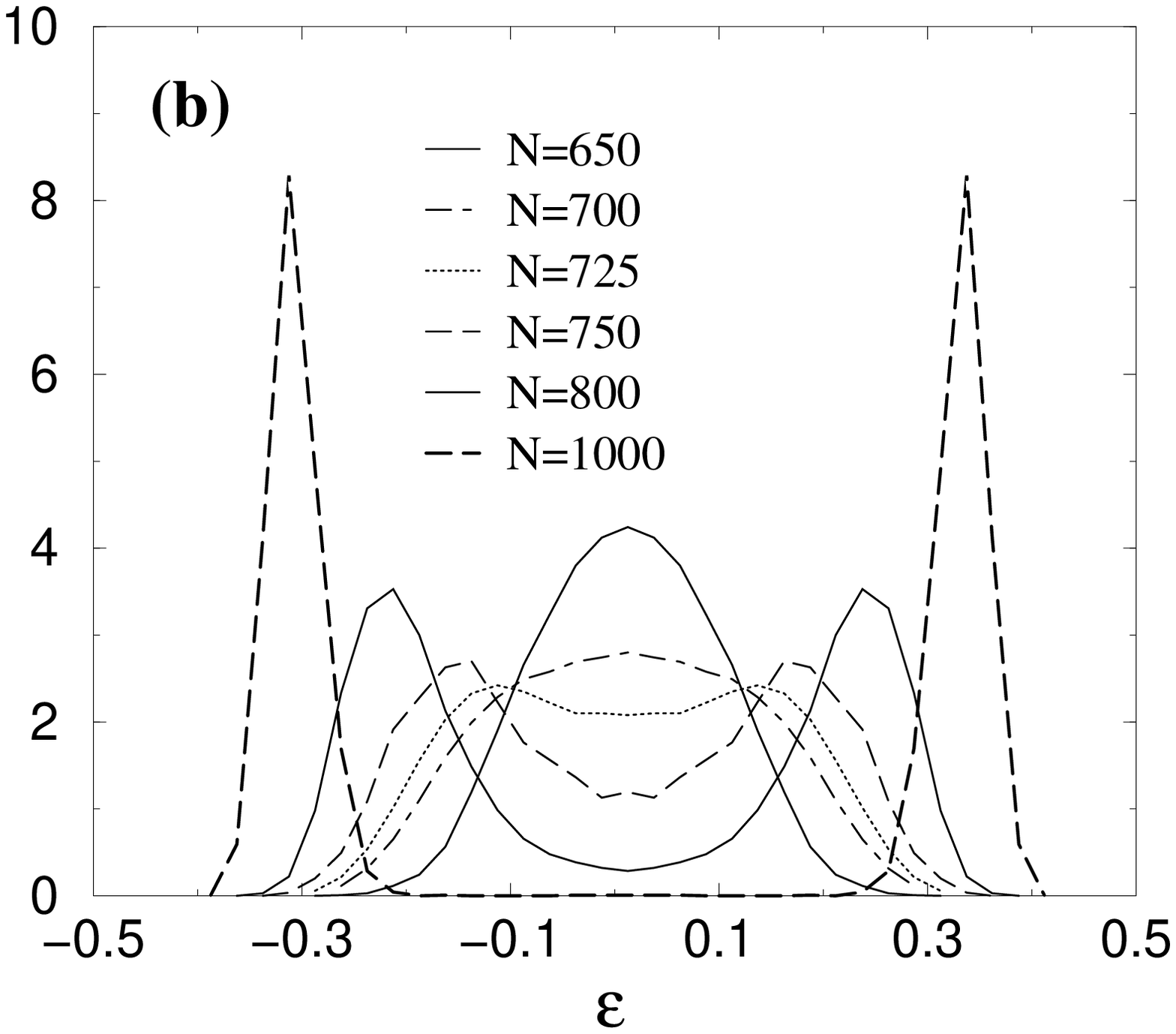,width=8cm,angle=0}
}
\caption{(a) Number of particles of type $2$ in the left compartment 
as a function of time (measured in number of collisions per particle),
for various values of $N$, and $m_1/m_2=3$. In all cases, the
horizontal lines correspond to the symmetric situation
$N^{left}=N^{right}$.\\ (b) Probability distribution function of
$\epsilon$ for $m_2/m_1=3$.  }
\label{fig:epsilon}
\end{figure}
\end{center}

However, for symmetry reasons, one should expect that the mean value
$\langle \epsilon\rangle$ (and the $\langle \epsilon_i\rangle$) always
vanish for sufficiently long simulation times, so that these
quantities are arbitrary and do not provide relevant order
parameters. Inspection of the time behaviour of $N^{left}$ confirms
this picture [see Fig.~\ref{fig:epsilon}-a)], which is made more
quantitative by computing the probability distribution function of
$\epsilon$ over very long runs and various initial conditions
[see Fig.~\ref{fig:epsilon}-b)].  At
small $N$, $\epsilon$ fluctuates around 0 and its standard deviation
increases with $N$. As $N$ increases, the asymmetric configurations
become stable but the system continuously jumps from one of the
possible asymmetric situations to the other, still spending some time
in between close to the symmetric state.  When $N$ further increases,
the residence time spent in each of the asymmetric states increases
and may eventually overcome the simulation time: for $N \gg N^*$,
starting from a symmetric situation, the system quickly evolves into
an asymmetric configuration, in which one compartment is strongly
overpopulated, and remains in this situation for all the simulation
time. For larger simulation times however the symmetry would be
restored. The situation is thus analogous to that of a two-states
system, in which the energy barrier between two symmetric states
increases with system size.  This behaviour is reminiscent of that
recently reported in~\cite{Meerson}: in this study of a translational
symmetry breaking as the aspect ratio of the simulation box is changed
(without a separating wall), large fluctuations have been shown to occur
in a wide region around a hydrodynamically predicted threshold value
beyond which the homogeneous system becomes unstable.

\begin{center}
\begin{figure}[ht]
\centerline{
\epsfig{figure=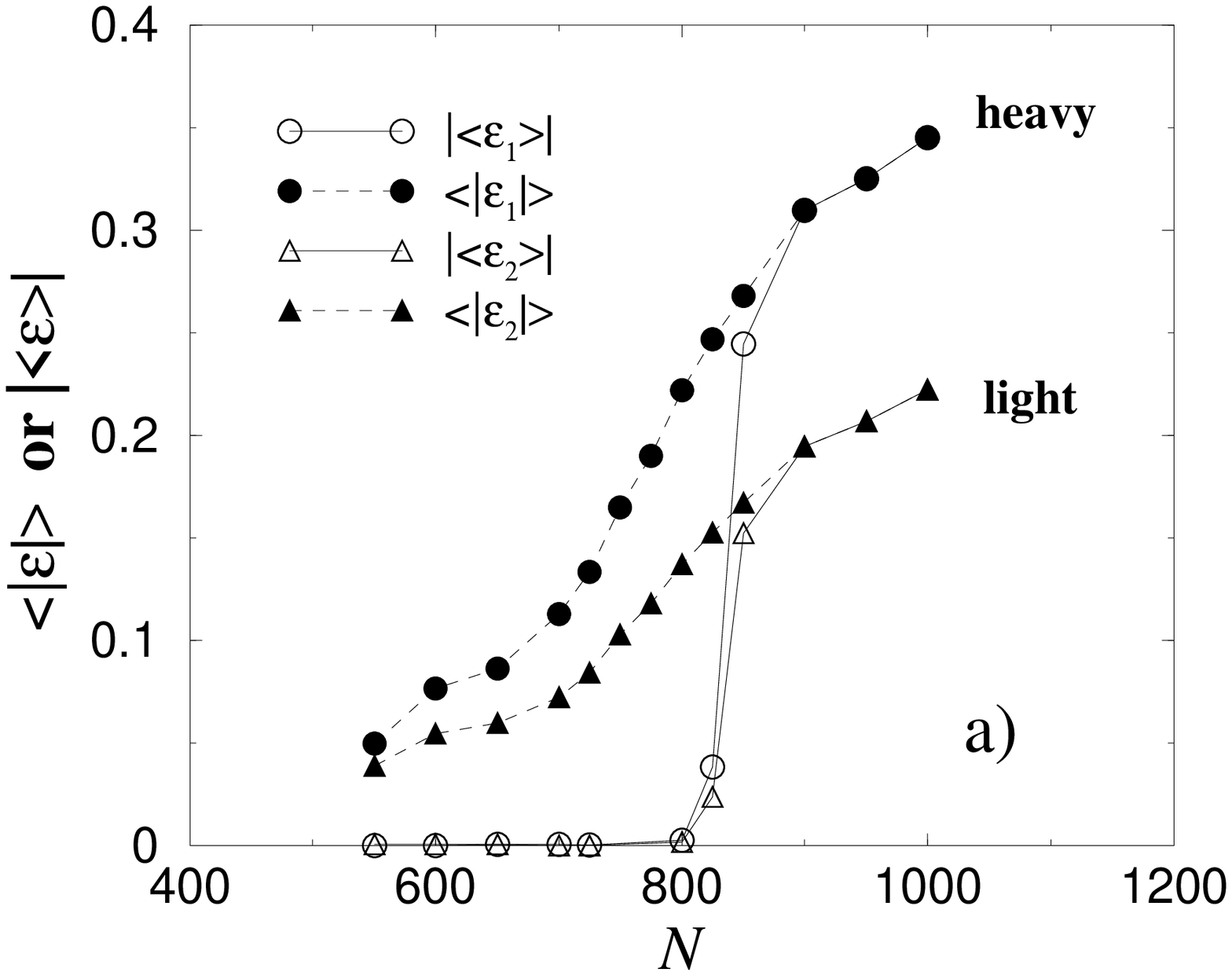,width=8cm,angle=0}
\epsfig{figure=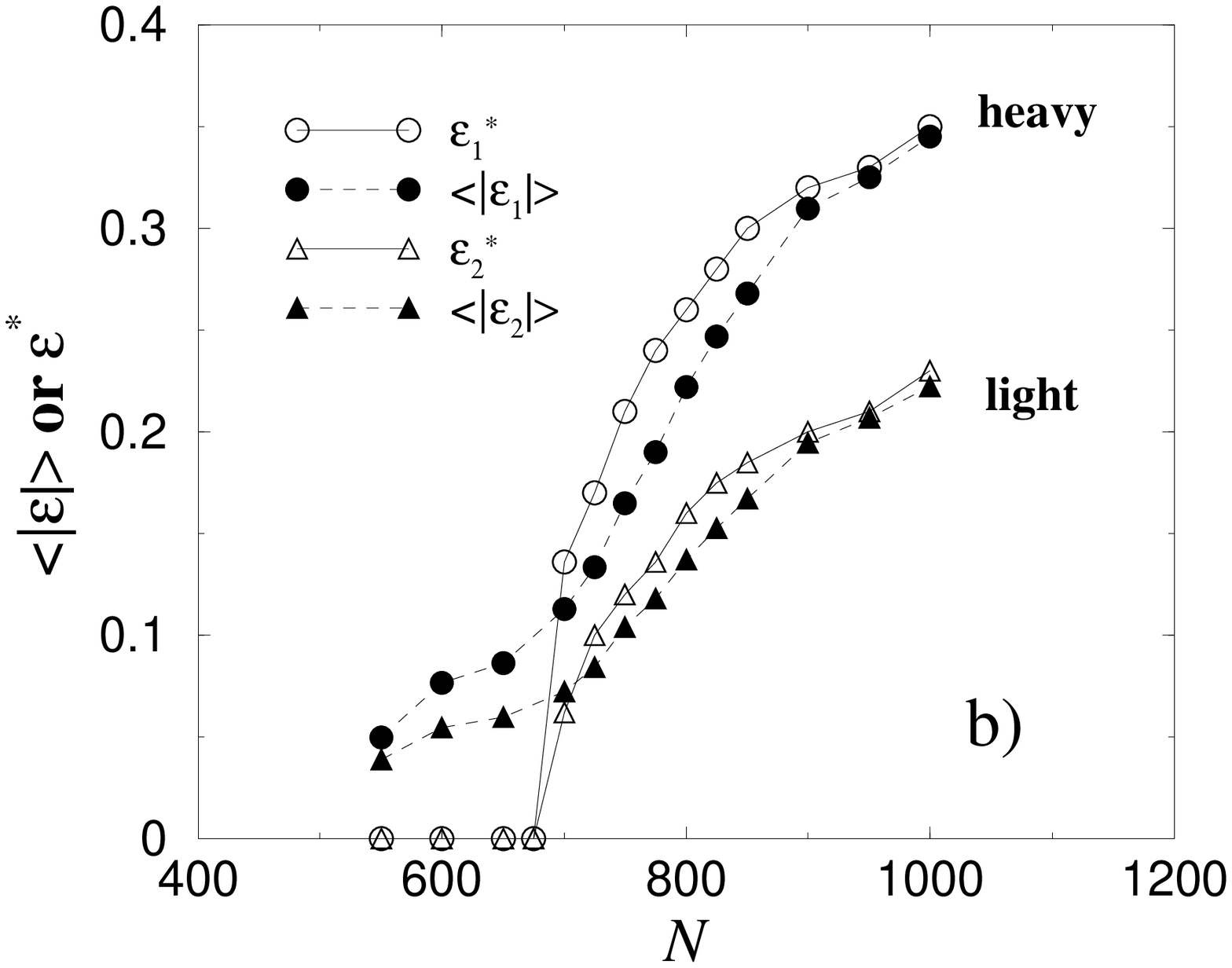,width=8cm,angle=0}
}
\caption{
(a) Asymmetry parameters $\langle|\epsilon_i|\rangle$ and 
$|\langle\epsilon_i\rangle|$ versus number
of particles for $m_1=5m_2$. The inelasticity 
coefficients are the same as in Fig.~\protect\ref{fig:diagram}
($\alpha_{ij}=0.9$).\\
(b) Comparison of the most probable values $\epsilon^*_i$ with
$\langle|\epsilon_i|\rangle$ for the same parameters as in (a).
}
\label{fig:abs}
\end{figure}
\end{center}

In any case, $|\langle\epsilon\rangle|$ vanishes for any $N$ for long
enough simulations, and does not provide an acceptable order
parameter. There are then {\it a priori} two possibilities to
construct such a quantity: a) by time averaging $|\epsilon|$ or b) by
extracting the most probable (say positive) value $\epsilon^*$ of
$\epsilon$ from its probability distribution function (p.d.f, see
Fig.~\ref{fig:epsilon}), averaged over the simulation time and over
various initial conditions.  Note that this p.d.f., and thus both
possible definitions, are not sensitive to the length of the
simulations (except for very small simulation times)~\cite{note}. 
We compare in
Fig.~\ref{fig:abs} these two definitions with the previous one, 
$|\langle\epsilon\rangle|$, computed again for a given (large)
simulation time. Since $\epsilon$ fluctuates around $0$ even at small $N$,
the $\langle|\epsilon_i|\rangle$ depend
rather smoothly on $N$, and therefore do not allow a clear definition
of a critical number of particles. On the other hand, the most
probable values $\epsilon_i^*$ allow to define a critical region,
being identically $0$ at small $N$ and taking positive values above a
certain threshold (see Fig.~\ref{fig:abs}). When $N$ is large enough,
the probability distribution functions of the $\epsilon_i$ become
sharply peaked around the $\epsilon_i^*$ so that both quantities
$\epsilon_i^*$ and $\langle|\epsilon_i|\rangle$ become close;
moreover, these p.d.f. take extremely small values in the vicinity of
$\epsilon_i=0$: this corresponds to the fact that the system is stuck
for long times in one of its two most probable states, so that the
computation of $|\langle\epsilon_i\rangle|$ coincides with that of
$\langle|\epsilon_i|\rangle$, or $\epsilon_i^*$.

\section{Density profiles and pressure tensor}
\label{sec:density}

\begin{center}
\begin{figure}[ht]
\epsfig{figure=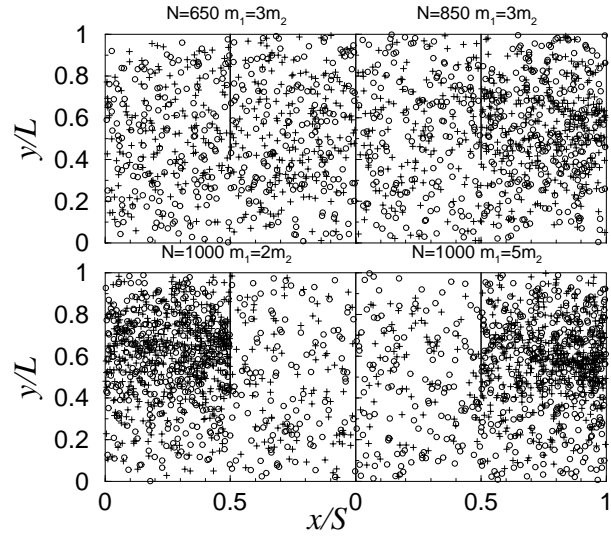,width=8cm,angle=0}
\caption{Typical instantaneous snapshots. Heavy particles (label $1$) 
are denoted by a plus, and light particles (label $2$) by a circle.
Top left: $N=650$, $m_1=3m_2$; Top right:
$N=850 $, $m_1=3m_2$; Bottom left $N=1000$, $m_1=2m_2$; Bottom right:
$N=1000$, $m_1=5m_2$. 
}
\label{fig:snapshots}
\end{figure}
\end{center}

\begin{center}
\begin{figure}[ht]
\centerline{
\epsfig{figure=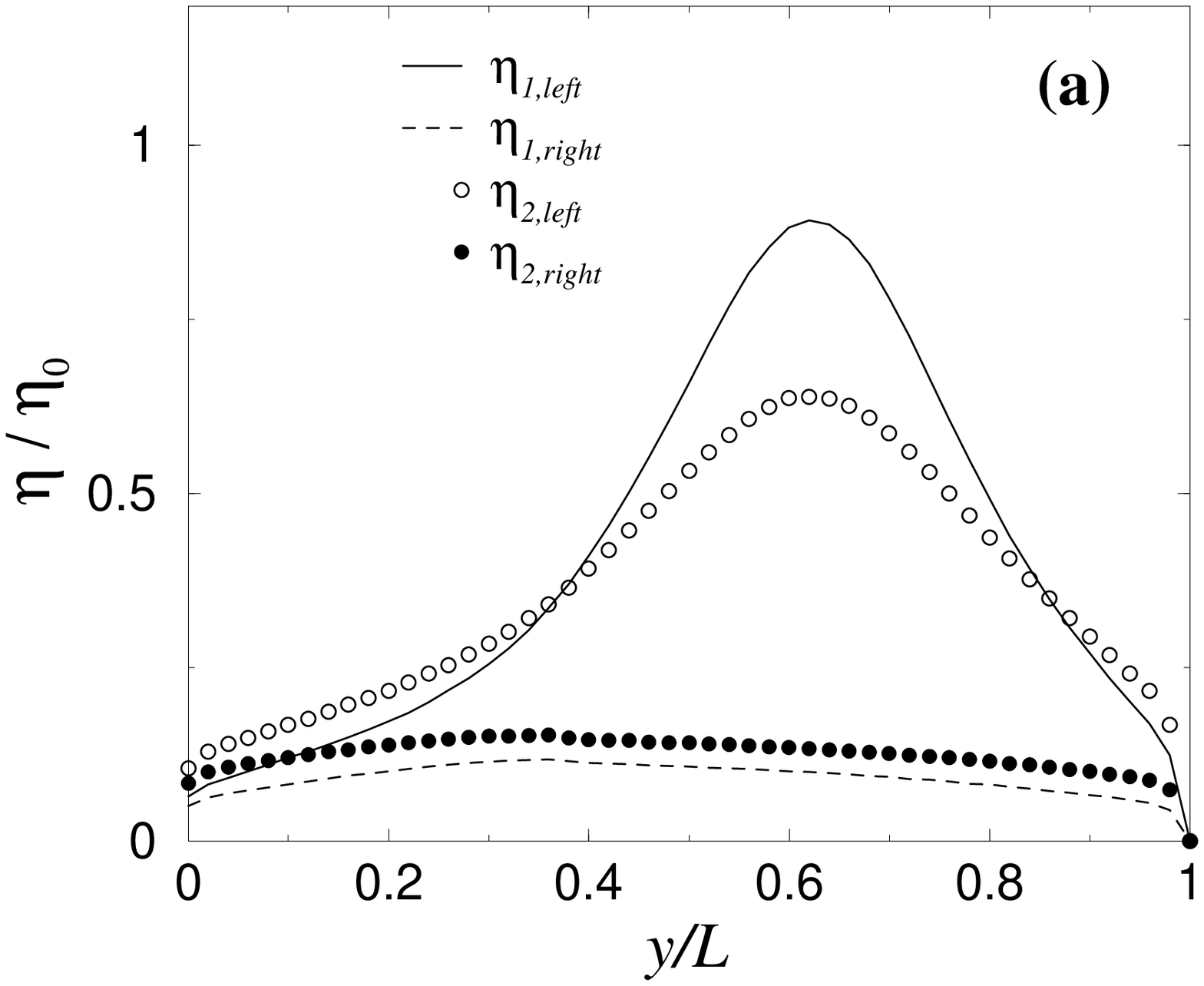,width=8cm,angle=0}
\epsfig{figure=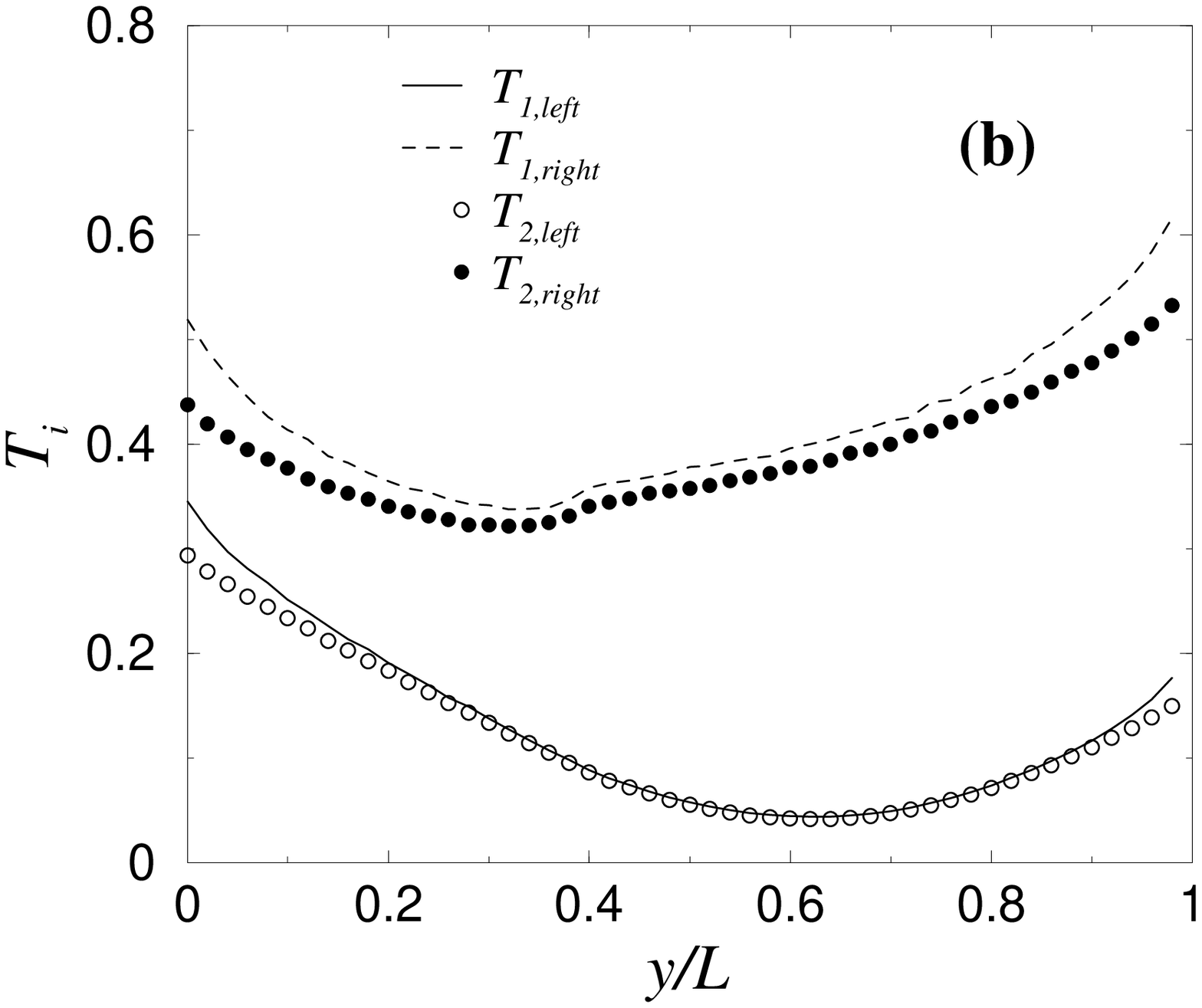,width=8cm,angle=0}
}
\caption{(a) Density and (b) temperature profiles for $N=1000$, 
$m_1=2m_2$. The left compartment ($0<x<S/2$) is denser and colder than
the right one. In the right compartment, the light particles are
denser than the heavy ones.  The mean packing fraction, averaged over
the whole system is $\eta_0=0.015$. The ratio $\eta(y)/\eta_0$ is also
the ratio $\rho(y)/\rho$ of local density normalized by the mean one.
The separation between the two compartments is located at $x=S/2;
0.4<y/L<1$.  }
\label{fig:profiles}
\end{figure}
\end{center}

Instantaneous typical configurations are displayed in
Fig.~\ref{fig:snapshots} for various values of $N$ and mass ratio
$m_1/m_2$.  From the coarse grained local packing fractions
$\eta_i(x,y)$, 
we define $x$-averaged quantities in each compartment:
\begin{equation}
\eta_i^{l} (y) = \frac{2}{S} \int_0^{S/2} dx \ \eta_i(x,y)\ \ , \ \ 
\eta_i^{r} (y) = \frac{2}{S} \int_{S/2}^S dx \ \eta_i(x,y)\ \ , i=1,2.
\end{equation}
These quantities are averaged over time for one run between two
successive ``flips'' (see section~\ref{sec:bifurc}), but not averaged
over various runs since the asymmetry would then be lost. The
corresponding density and temperature profiles are shown in
Fig.~\ref{fig:profiles} for $N=1000$, well above the bifurcation
point.  One may observe that in the asymmetric situation, the
densities are different even for $y < y_0$, i.e. not only where the
compartments are physically separated.

\begin{center}
\begin{figure}[ht]
\centerline{
\epsfig{figure=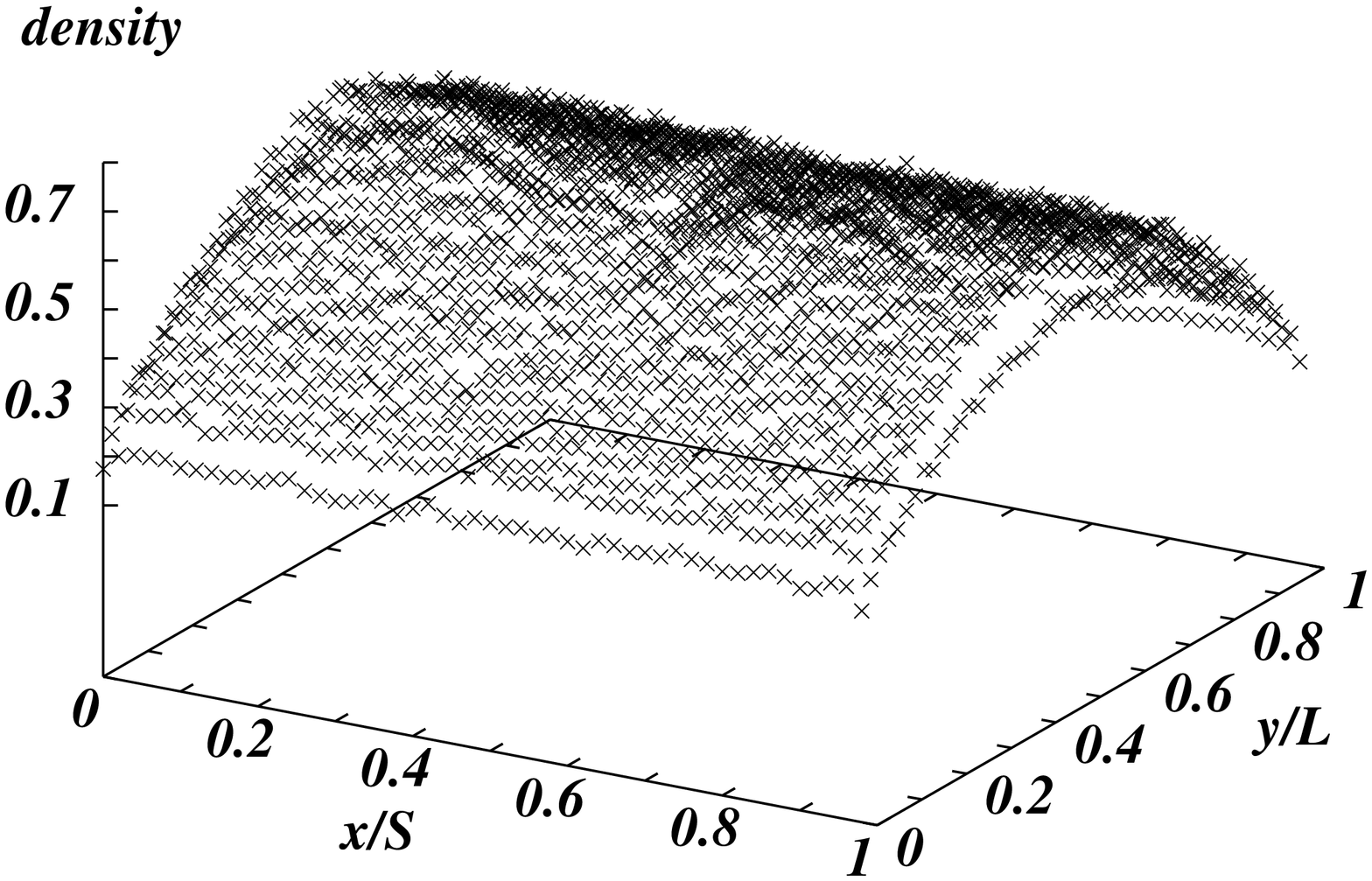,width=5cm,angle=-90}
\epsfig{figure=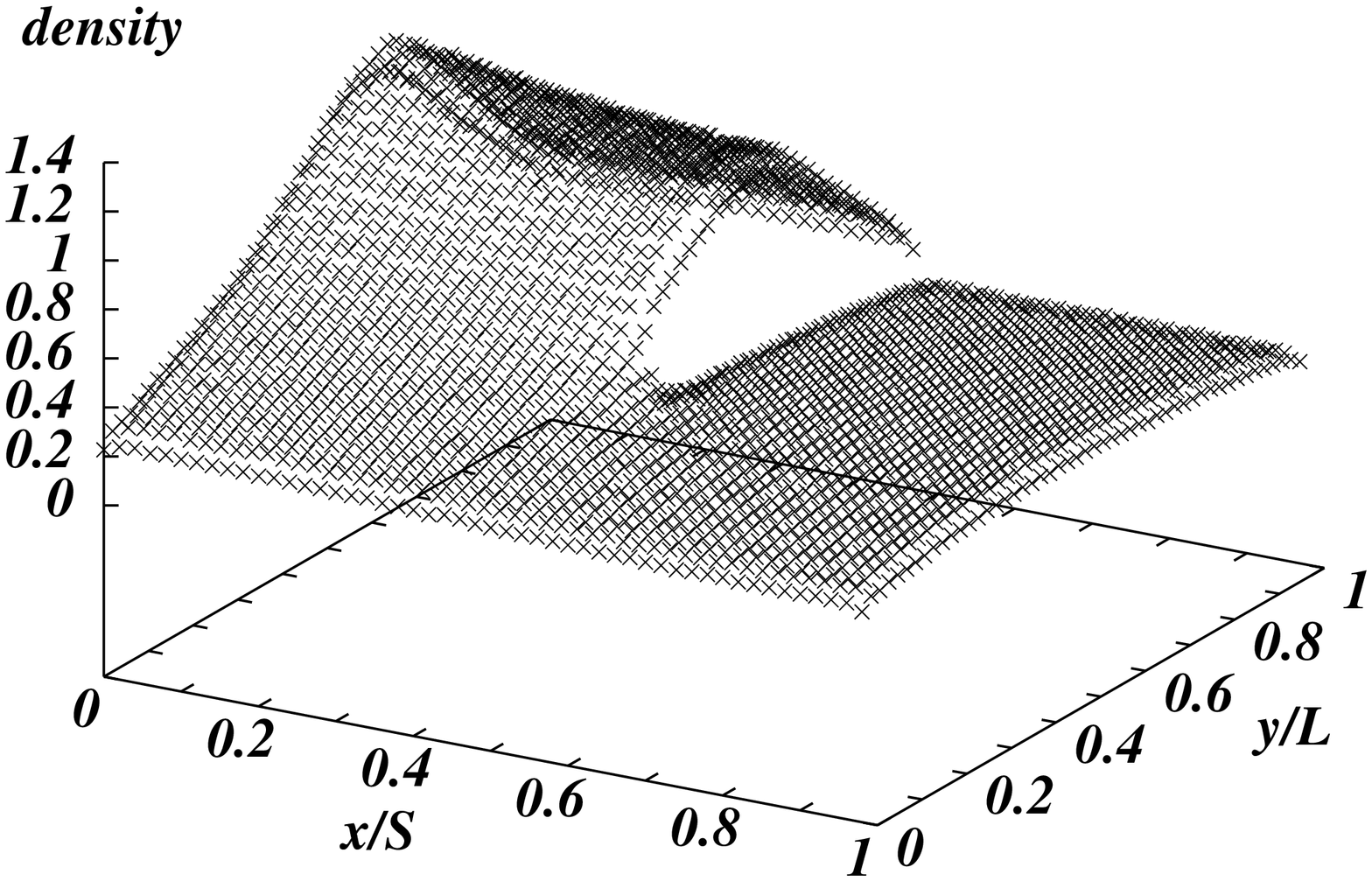,width=5cm,angle=-90}
}
\caption{Averaged local density of the particles of
type $2$ (light component) for $N=400$ (left panel) and 
$N=900$ (right panel). 
The separating wall is at $x=S/2$, $y > y_0=0.4L$
and  $m_1=3m_2$ in both cases.
}
\label{fig:2dplotdens}
\end{figure}
\end{center}

Two-dimensional plots of the coarse grained densities $\eta_i(x,y)$
are displayed in Fig.~\ref{fig:2dplotdens} for two values of the
number of particles, well below and well above the bifurcation.  Below
the transition, translational invariance in $x$ holds in the whole
box, while above, the densities and temperatures are almost
independent of $x$ in each compartment, but are discontinuous at
$x=S/2$ for $y > y_0$ because of the separating wall; at $y < y_0$ but
close to $y_0$ a quite sharp change is observed in the vicinity of
$x=S/2$.  At small $y$ the $x$-gradients are smaller.


In the hydrodynamic study of~\cite{Brey_Maxwell}, the ideal gas form
for the pressure constitutes a fundamental hypothesis which allows for
an analytic treatment; moreover, the pressure is assumed to be
isotropic. However, anisotropic energy injection mechanisms lead to
anisotropic pressure tensors, especially near vibrating
walls~\cite{Brey_aniso,vibrated}. Knowing the density and pressure
profiles for our system, one may compute the two components $P_{xx}$
and $P_{yy}$ of the pressure tensor from a given equation of state. We
consider the generic expression derived in~\cite{vibrated} within
Enskog-Boltzmann kinetic theory. For an homogeneous and isotropic
mixture with partial temperatures $T_i$, number fractions $x_i$, and
without any approximation on the single particle velocity
distribution, it was obtained that
\begin{equation}
P \,=\, \sum_{i} \rho_i T_i \,+\, \rho \eta\, 2^{d-1}\,\sum_{i,j} 
x_ix_j\, \frac{m_j}{m_i+m_j} \, (1+\alpha_{ij}) \,\, T_i \, \,
\frac{\sigma_{ij}^d}{\langle \sigma^d\rangle} \, \, \chi_{ij},
\label{eq:eos}
\end{equation}
where $d$ denotes space dimension, $\sigma_{ij} =
(\sigma_i+\sigma_j)/2$, $\langle\sigma^d\rangle = \sum_{i} x_i
\sigma_i^d$. The $\chi_{ij}$ are the {\it a priori} unknown pair
distribution functions at contact; these quantities embody the
correction to the ideal gas equation of state, and since we are
considering dilute system, it is sufficiently accurate to assume
$\chi_{ij}=1$ (low density limiting value). The values of $P_{xx}$ and
$P_{yy}$ are finally obtained by substituting $T_i$ in (\ref{eq:eos})
respectively by $T_{ix}$ and $T_{iy}$ (i.e. the mean square $x$ or $y$
components of particles velocities).

\begin{center}
\begin{figure}[ht]
\centerline{
\epsfig{figure=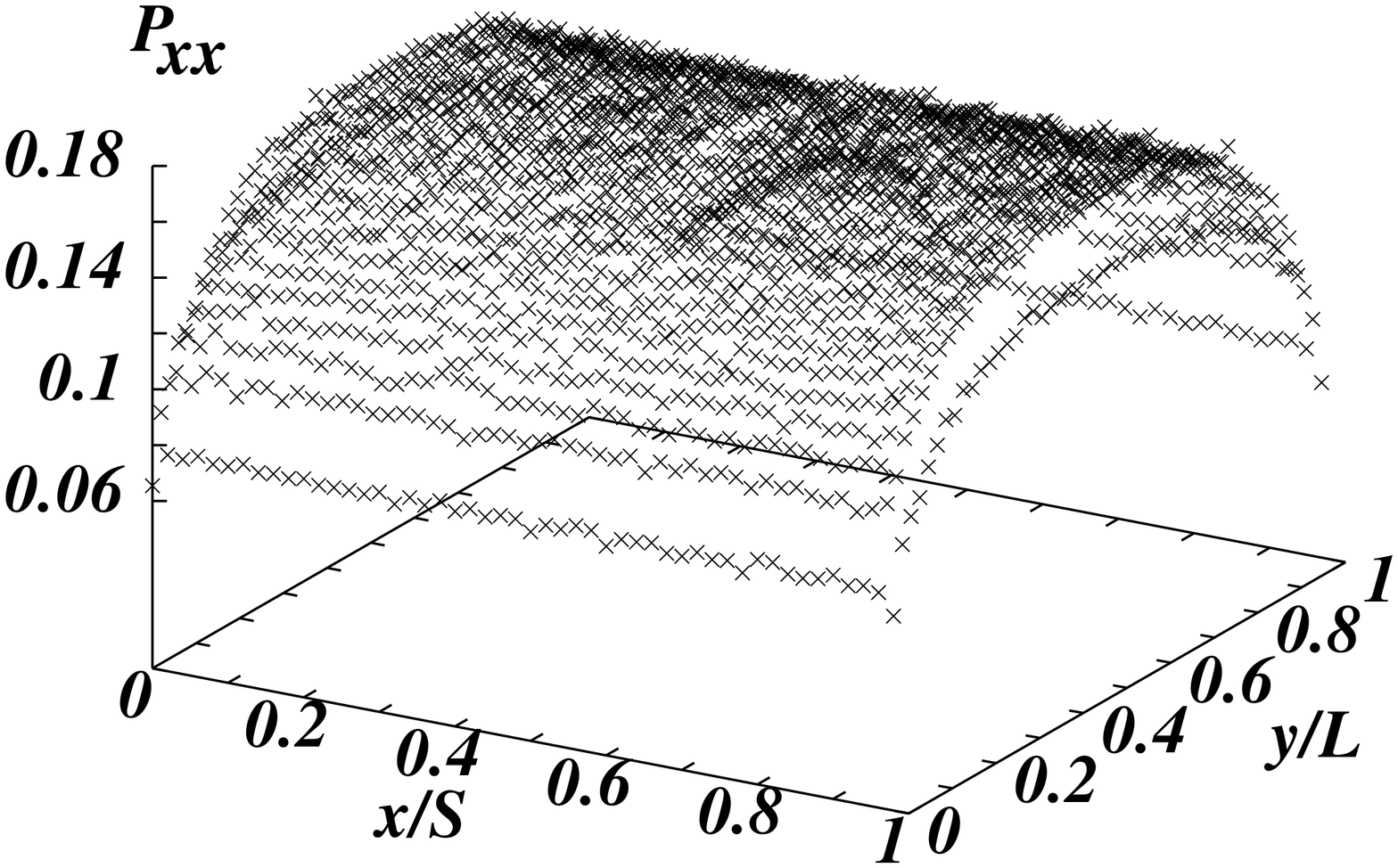,width=4cm,angle=-90}
\epsfig{figure=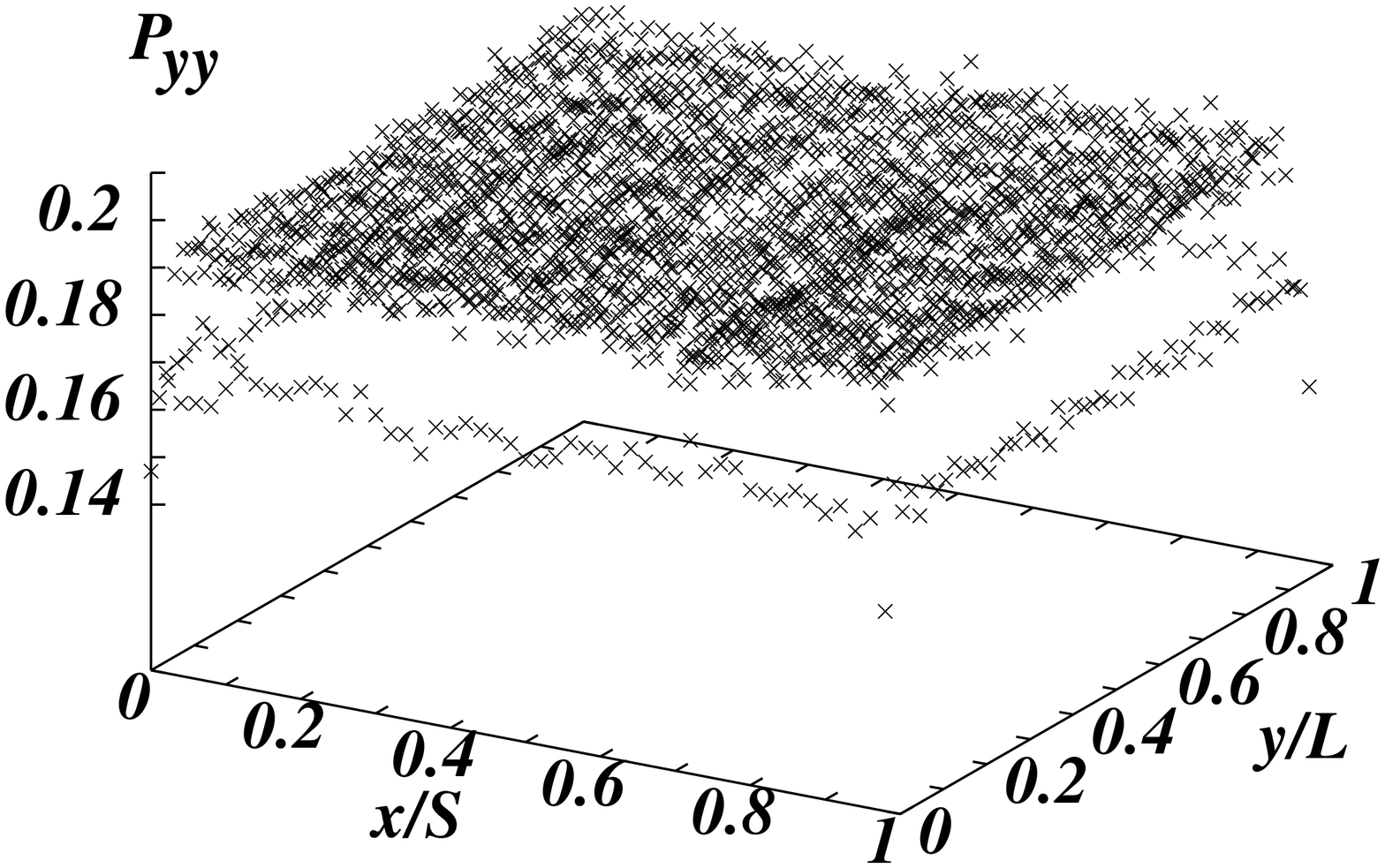,width=4cm,angle=-90}
}
\centerline{
\epsfig{figure=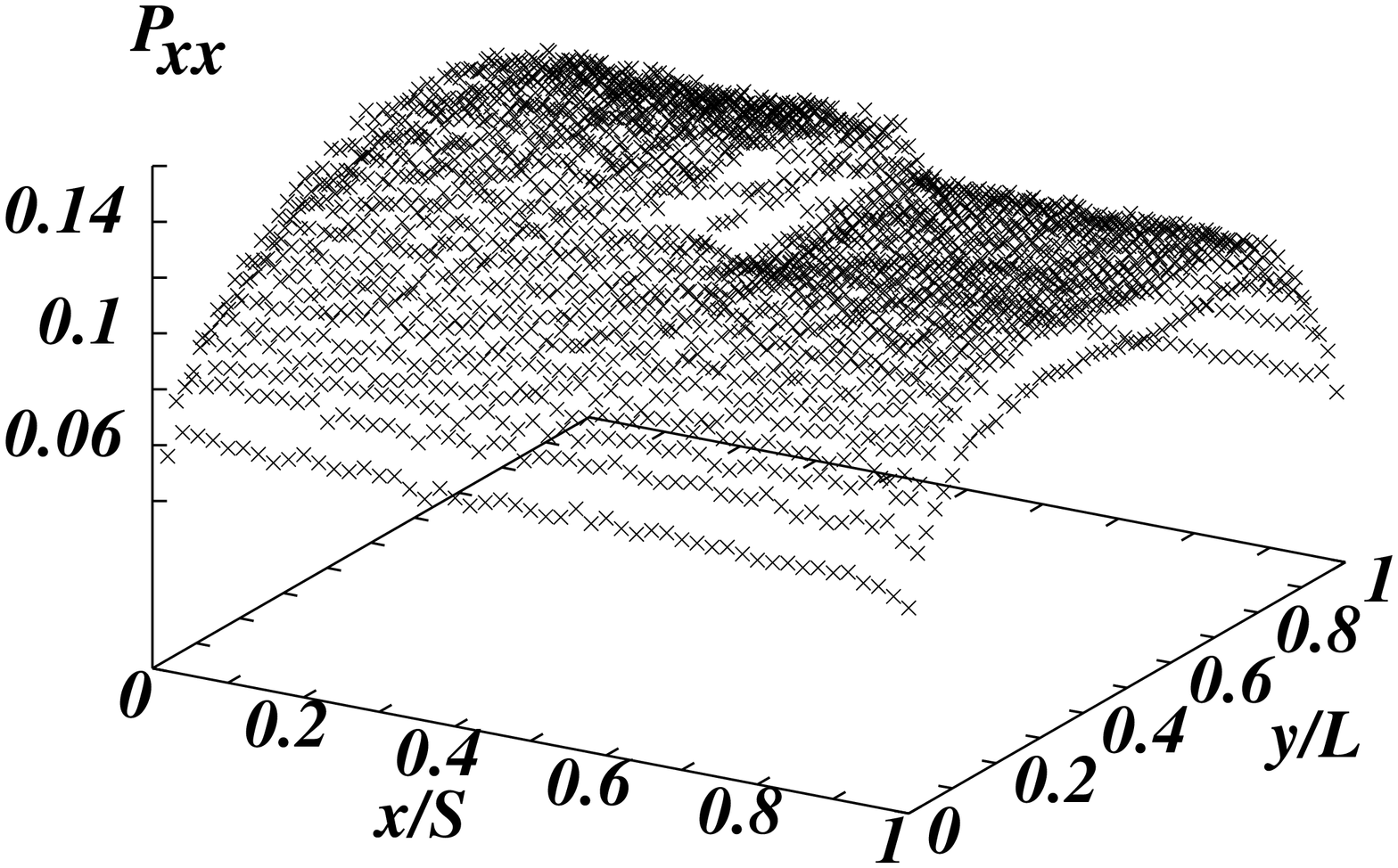,width=4cm,angle=-90}
\epsfig{figure=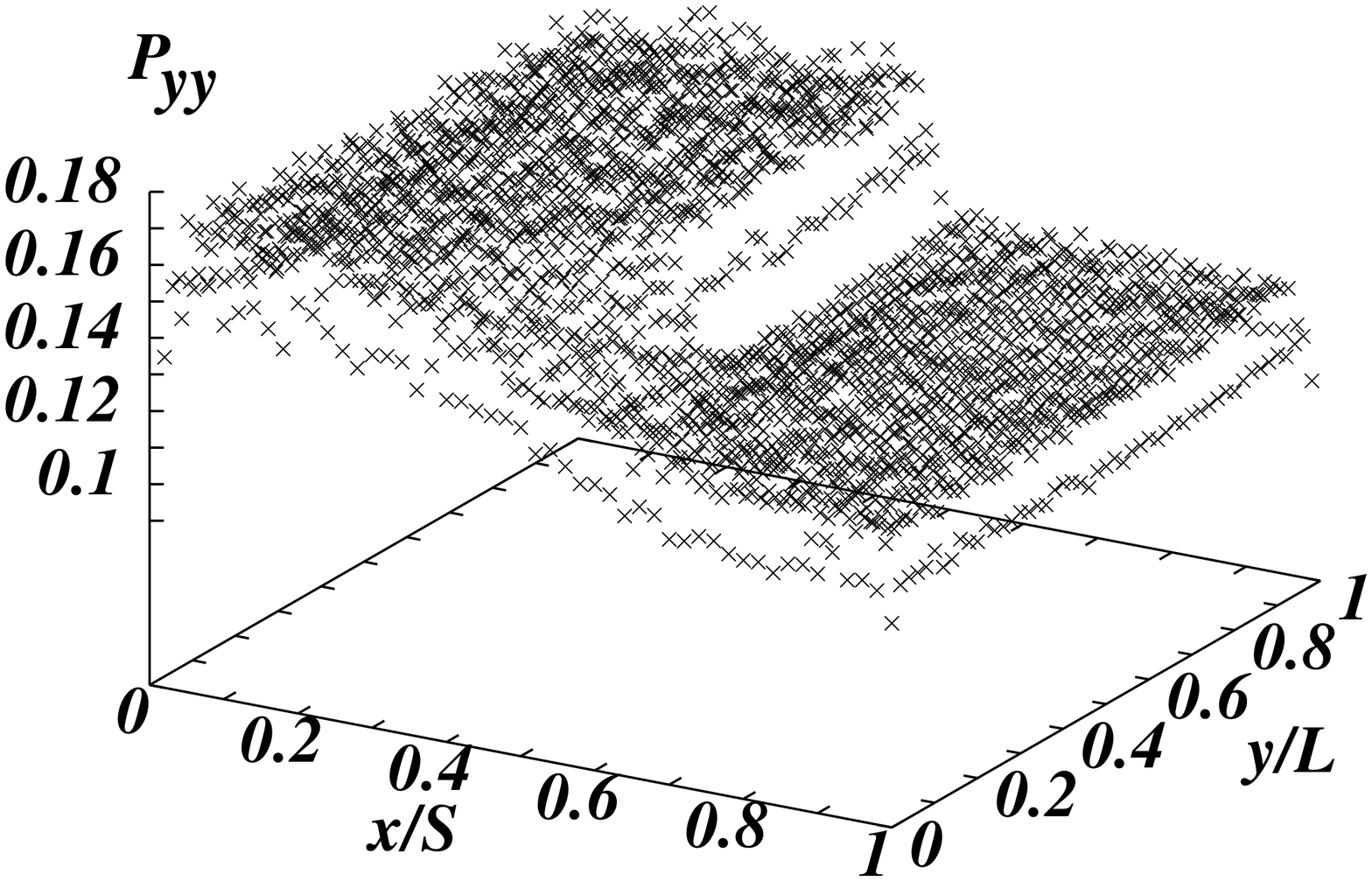,width=4cm,angle=-90}
}
\caption{Components of the pressure tensor  $P_{xx}$ (left)
and $P_{yy}$ (right), as given by the equation of state [Eq. (\ref{eq:eos})], 
for $N=600$ (top) and $N=900$ (bottom). Here, $m_1/m_2=3$ and 
$\alpha_{ij}=0.9$. For $N=900$, the right compartment ($x > S/2$) is
more populated, colder and at a lower pressure.
}
\label{fig:pressure}
\end{figure}
\end{center}

The results are summarized in Fig.~\ref{fig:pressure}, where we plot
$P_{xx}(x,y)$ and $P_{yy}(x,y)$ for $N=600$ and $N=900$.  Below the
transition ($N=600$), the picture is similar to the one without a
separating wall, and the pressure tensor is $x$ independent. One
therefore has $\partial_x P_{xx}=\partial_x P_{yy}=\partial_y
P_{yy}=0$.  However, for $N=900$ (above the transition), the $yy$
components are no longer equal in the left and right sides, while the
$xx$ components are equal only for $y < y_0$ where the separation
begins.  One still observes $\partial_x P_{xx}=\partial_y P_{yy}=0$ in
each compartment, except close to the extremity of the separating wall
($x=S/2,y=y_0$), but for $y>y_0$ the separating wall allows for
different values of the pressure components. Moreover $\partial_x
P_{yy} \ne 0$ for $y<y_0$ while for $y> y_0$, $\partial_x P_{yy} = 0$
in each compartment (except close to $x=S/2$), with a discontinuity at
$x=S/2$. It is noteworthy that the denser compartment is also the one
where both components of the pressure tensor are lower, since it is
much ``colder'' than the dilute compartment.

This analysis shows that both above and below the symmetry breaking,
the (anisotropic) pressure tensor as computed from Eq. (\ref{eq:eos})
is divergence free: $\partial_x P_{xx}+\partial_y P_{yy}=0$. This
``hydrostatic'' requirement follows from the condition of a vanishing
flow field, and in spite of the low mean densities considered here,
would not be fulfilled restricting $P_{xx}$ and $P_{yy}$ to their
ideal parts.

\section{Conclusions}
\label{sec:concl}

For a one component granular gas enclosed in a box made of two
connected compartments, a vigorous shaking is known to promote a
symmetry breakdown and separate the system into a cold and dense
region on the one hand, and a hot and dilute part on the other hand
(so called ``Maxwell-Demon'' experiment).  In addition, in a binary
granular mixture, heavy and light grains generically have different
granular temperatures. In this contribution, we have combined both
aspects (Maxwell Demon and mixture) to investigate the possibility to
separate the two components of the mixture.  Our molecular dynamics
results show a spontaneous symmetry breaking as the number of
particles is increased, all other parameters being kept constant. The
denser compartment then appears to be rich in heavy particles, but
this partial segregation is such that this compartment is also richer
in light particles than the other half of the confining box (which is
however the light-rich one). It therefore seems that such a set-up
cannot achieve an efficient segregation (although a possibility would
be to isolate the dense compartment and iterate the process with this
non-equimolar mixture).

The transition reported here is not {\it stricto sensu} a phase
transition, since the control parameter is the system size.  As a
consequence, fluctuations can always bring the system from one of the
asymmetric states to the other, as e.g. for a finite size Ising model
below its critical temperature. We have discussed the consequences of
this feature on the definition of a relevant order parameter to
characterize the bifurcation.

We finally note that the experimental realization of the two
dimensional situation investigated here seems feasible, for instance
by adapting the configuration used in~\cite{Rouyer} (friction with the
walls confining the system in a 2D slab might play an role, and has
not been considered here). The experimental signature of the large
fluctuations that invalidate hydrodynamic approaches seems an
interesting issue.


\null
\vskip 3cm 
\begin{center}
\begin{figure}[ht]
\centerline{
\epsfig{figure=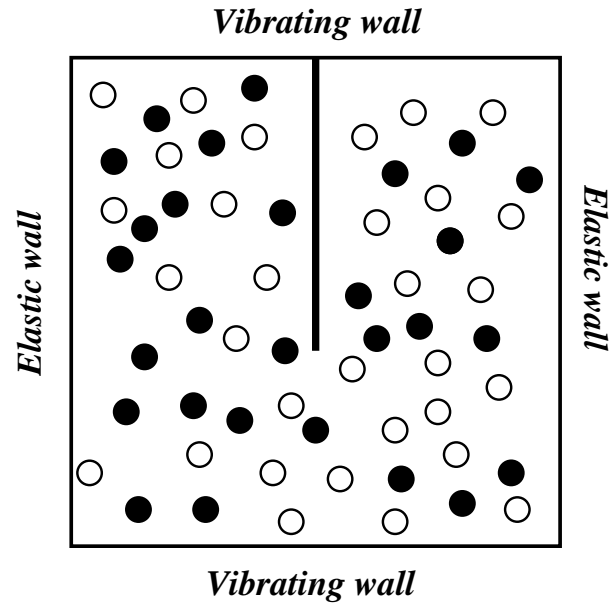
,width=8cm,angle=0}
}
\caption{Schematic picture of the set-up}
\label{fig:setup}
\end{figure}
\end{center}

\bigskip
\null\vskip 2cm
\bigskip

\begin{thebibliography}{99}
\bibitem{Rqe}
In the context of homogeneously heated binary granular mixtures, 
it has been shown in \protect\cite{Granular} that the influence of size
asymmetry on the kinetic energy non-equipartition is much less important
than that of mass asymmetry. We therefore restrict our study to the
influence of the latter parameter.
\bibitem{note}
Note that similar definitions could also be used in the context
of~\protect\cite{Meerson} to construct order parameters.
\end{thebibliography}
\end{document}